\documentclass[letter,twocolumn]{jpsj2} 

\title{Molecular-Dynamics Simulation of Vulcanian Eruption}

\author{Satoshi \textsc{Yukawa}\thanks{E-mail address:
    yukawa@ap.t.u-tokyo.ac.jp} 
and Nobuyasu \textsc{Ito}\thanks{E-mail address: ito@ap.t.u-tokyo.ac.jp}}

\inst{
Department of Applied Physics, School of Engineering, 
The University of Tokyo,\\7-3-1, Hongo, Bunkyo-ku, 113-8656, JAPAN
}

\abst{
Vulcanian explosive eruption, 
which is a nonlinear and nonequilibrium abrupt dynamics of magma-gas mixture, 
is modeled by a two-component Lennard-Jones particle system. 
Molecular-dynamics simulation of a shock-tube experiment 
gives consistent results with a explosive eruption picture of volcanology; Shock wave and expansion wave are reproduced. In addition bubble nucleation of a gas component in the magma melt and spinodal-like decomposition are observed in the simulation. 
The result is also compared with a continuum hydrodynamic model;
Qualitative features of continuum dynamics are reproduced by the present model.
We find that 
the particle description of dynamics 
is an effective method in such kind of abrupt dynamics.
}

\kword{Vulcanian eruption, Lennard-Jones particle, molecular dynamics simulation, shock tube}

\begin{document}
\maketitle

Volcanic eruption is complicated physical phenomena and the physical
understanding has not been well established yet; The problem is to
understand nonlinear and nonequilibrium dynamics of magma-gas mixture
accompanied by phase transitions.\cite{MS99,Me00,Ko05} Existence of gas,  
which is mainly $\mathrm{H_2O}$, is sometimes
forgotten, but it is pointed out that such gas component plays an important role in explosive eruption.\cite{Sch04}
Type of volcanic eruption is classified into three classes by
chronological behavior; 
One is so-called Vulcanian type eruption, which  is widely observed in Japanese volcanos. 
This type is
characterized by an intermittent explosive eruption and formation of a
lava dome. 
These features are determined by physical properties of magma; 
Specifically viscosity of magma controls them. 

In this paper, we study Vulcanian eruption, because its explosive
mechanism 
will be the most interesting physically, 
in particular, in the context of nonequilibrium physics; 
In the volcanology, an eruption
picture is considered as follows: A stage of eruption dynamics
consists of a magma chamber and a conduit. Top of conduit is covered
by a lava dome.  In a top of magma chamber or a lower part of conduit,
a gas component is almost completely dissolved into the magma
melt. 
In the upper region of
saturated magma, the gas is exsolved according to the equilibrium
solubility law.  As decreasing the lithostatic pressure, volume
fraction of gases is increasing. 
At the beginning of eruption, pressure of the magma-gas mixture
is considered to 
increase, although the mechanism 
is not clear yet. 
When the lava dome cannot support this overpressure, it disrupts the
lava dome.  At the next moment, two shock waves appear and propagate;
One is a  shock wave formed between atmosphere and compressed air and it propagates upward. 
Another is decompression wave in magma-gas melt and it goes to
opposite direction.  During the eruption it is observed that the
transition from the laminar flow of bubbly melt to the turbulent flow
of gas-magma dispersion in the conduit.  This transition layer
determines the front of fragmentation wave which propagates downward. 
At the moment, viscosity of magma-gas mixture is drastically
changed abruptly about the order of $10^{12}\sim 10^{15} \mathrm{Pa\cdot s}$.
%
%

There have been many theoretical investigations of Vulcanian eruption
in the volcanism study.  In 1995, Woods proposed the model for magma
flow in conduit;\cite{Wo95} In his model magma-gas mixture is treated as a
one-dimensional nonviscotic compressible fluid with single component.
This model can capture physical properties of dynamics in some sense.
But treatment of dynamics is not well satisfied; 
For example, flow is treated as isentropic one, though
bubble nucleation accompanies the eruption. 
There are some other phenomenological models, but 
the present understanding of the eruption dynamics is still unsatisfactory 
in the context of nonequilibrium physics.\cite{Me00,Ko05}

Recent progress of experimental techniques enables us to compare such
theoretical model with experimental results; These experiments are
called as shock-tube experiment.\cite{ZSS97,CBMS02,SDA04,IRS02} In the experiment, analogue materials of
magma-gas mixture, such as viscoelastic materials and powder, are used. 
It is observed that the behavior of  explosion depends on 
the viscosity of analogue materials. 
Thus a non-viscotic  treatment in a
theoretical study is not sufficient.

In this paper, we try to establish a computational microscopic model of Vulcanian eruption; 
So to say, we want to make ``an Ising model of Vulcanian eruption''.
Here we describe dynamics of the mixture by microscopic particle dynamics.
A particle dynamics simulation can be regarded as an ideal 
shock-tube experiment, because  we can calculate 
macroscopic quantities.  
In addition, using the particle
dynamics, we can also reproduce hydrodynamic behavior described by a
continuum description of Navier-Stokes equation.  Even in
Newtonian dynamics, we can produce macroscopic behavior in linear
nonequilibrium thermodynamic regime.\cite{IMYI04,MSIY03,OI03,OH04}
Moreover we can also discuss phenomena in far from equilibrium state, 
which are not captured by continuum descriptions based on local equilibrium.
Thus the particle model enable us to explore nonequilibrium 
dynamics of volcano, as well as the model 
can verify an macroscopic theoretical model.



Here we assume microscopic dynamics are governed by the following
Hamiltonian:
\begin{equation}
\mathcal{H} = \sum_{i=1}^N \dfrac{\mathbf{p}_i^2}{2m_i} 
+ \dfrac{1}{2} \sum_{i,j}^N \alpha_i \alpha_j \phi(\lvert \mathbf{q}_i
- \mathbf{q}_j \rvert) \enspace ,
\end{equation}
where $\phi(r)$ is Lennard-Jones 12-6 potential:
$ \phi(r) = 4 \epsilon \{
\left(\sigma/r\right)^{12} - \left(\sigma/r\right)^6\} 
+\phi_0
$. 
For computational efficiency, we introduce a potential cutoff as $3.9 \sigma$ and 
determine the value of $\phi_0$ to be $\phi(3.9 \sigma ) = 0$.  
And $N$ denotes total particle number, $m_i$ denotes mass of particle $i$,
$\mathbf{p}_i$ and $\mathbf{q}_i$ denote particle three-dimensional momenta
and coordinates, respectively.  
Dimensionless parameters $\alpha_i$ and 
$m_{\mathrm{gas}}/m_{\mathrm{magma}}$
are 
selected 
so that it will reproduce similar properties as 
magma gas;\cite{Sch04}
We take $\alpha_i$ to be $1$ for magma particles, and $0.1$ for gas
particles. 
It determines energy scales of magma and gas. 
Ratio of melting temperatures of magma to gas is given by 
$\alpha_{\mathrm{magma}}^2/\alpha_{\mathrm{gas}}^2 $ and it is $100$ in the present model, although 
it is approximately $1000$ for actual magma and gas. 
Present choice is ten times less than actual situation, but it is sufficient to 
describe the explosive eruption as we will show in the following. 
Particle mass ratio is chosen as $m_{\mathrm{gas}} / m_{\mathrm{magma}} =
0.1$, which is of order actual mass ratio.
Hereafter we measure length, mass, and energy by the units of 
$\sigma, m_{\mathrm{magma}}$ and $\epsilon$, respectively, and use
dimensionless variables. 
Employing the Lennard-Jones 12-6 potential makes us to describe
thermodynamic phases of  
gas, fluid, solid, and their coexisting state.

\begin{figure}[t]
  \centering
  \includegraphics[width=0.43\textwidth]{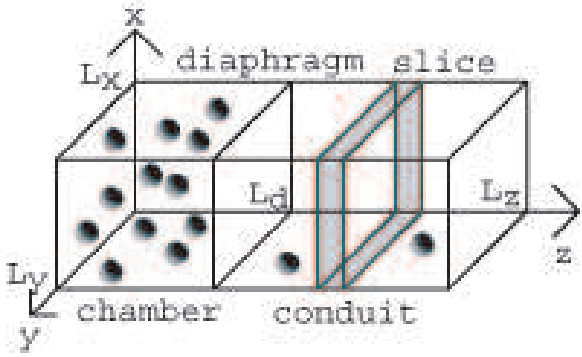} 
 \caption{Geometry of the system. When we calculate physical quantities, we slice the system with a unit length.}
  \label{fig:geom}
\end{figure}
Using the above Hamiltonian, we calculate particle motion.
The geometry of the system is as follows (see also Fig.~\ref{fig:geom}):
Consider rectangular parallelepiped with a size $L_x \times L_y \times L_z$.
For $x$ and $y$ directions, periodic boundary conditions are imposed. 
A eruption direction is to $z$ axis, and we prepare  elastic walls at
bottom and top.  These walls are represented by repulsion part of
Lennard-Jones potential. 

First we have to prepare initial state as thermal equilibrium
one. In this stage, whole system is divided into two parts, ``chamber'' ($0 \le z \le L_d$) 
and ``conduit'' ($L_d \le z \le L_z$) by a diaphragm, which is located at $z=L_d$, made of 
same elastic walls at $z=0$ and $z=L_z$.  At the beginning, magma and gas
particles are contained in the chamber. Contrarily, only gas particles
are in the conduit.  For preparing initial state, we do an isothermal
simulation with No\'{s}e-Hoover thermostat in each part of the system.\cite{No84-1,No84-2,Ho85}
Density and temperature in the chamber are chosen as gas particles are uniformly mixed into magma particles; There is no phase separation.

After thermalization, we remove the separator between conduit and
chamber and we detach the thermostat. Then the system  obeys the
Hamiltonian dynamics. If pressure in the chamber is higher than one in 
the conduit, an explosion is activated.

Simulation details are as follows: 
The second order symplectic method (the leapfrog method) is
used in numerical integration. Time  integration slice is  taken to be
$10^{-3}$. This value is sufficient for present simulations, which is checked by energy conservation.


In the simulation, we calculate several physical quantities in boxes
which are obtained by slicing along $z$-direction with a unit length
$\sigma$.\cite{IK50,HM86} 
Number density $n(z)$ and mass density $\rho(z)$ of the
slice $z$ are basic quantities of macroscopic dynamics defined by counting a number and mass in the local slice. Barycentric velocity $\mathbf{v}(z)$ is defined through sum of momenta in the slice. Pressure $p(z)$, is defined by a trace of stress tensor. 
And temperature $T(z)$ is defined by variance of particle velocities from local barycentric motion.

\begin{figure}[t]
  \centering
  \includegraphics[width=0.23\textwidth]{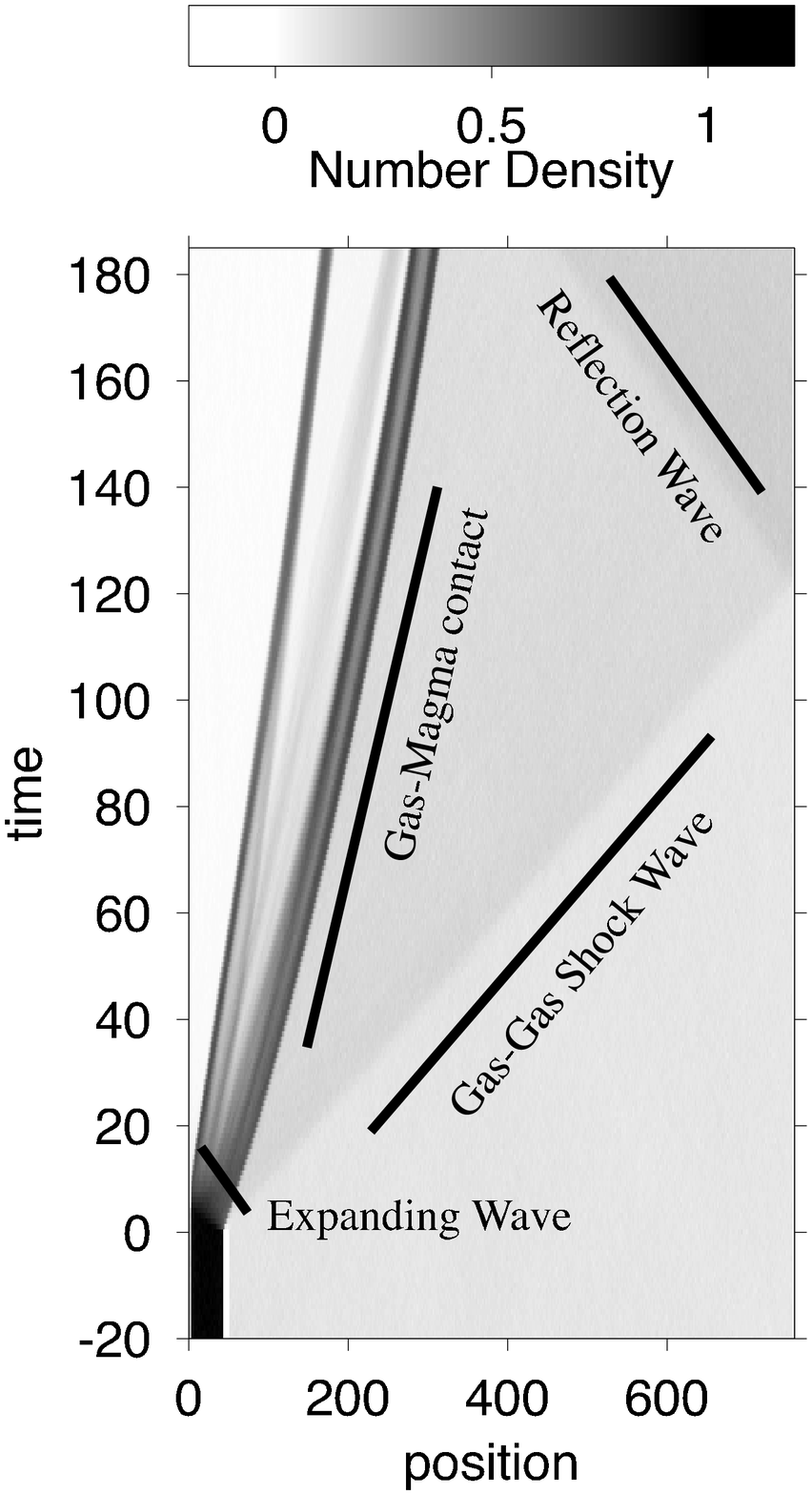}
  \includegraphics[width=0.23\textwidth]{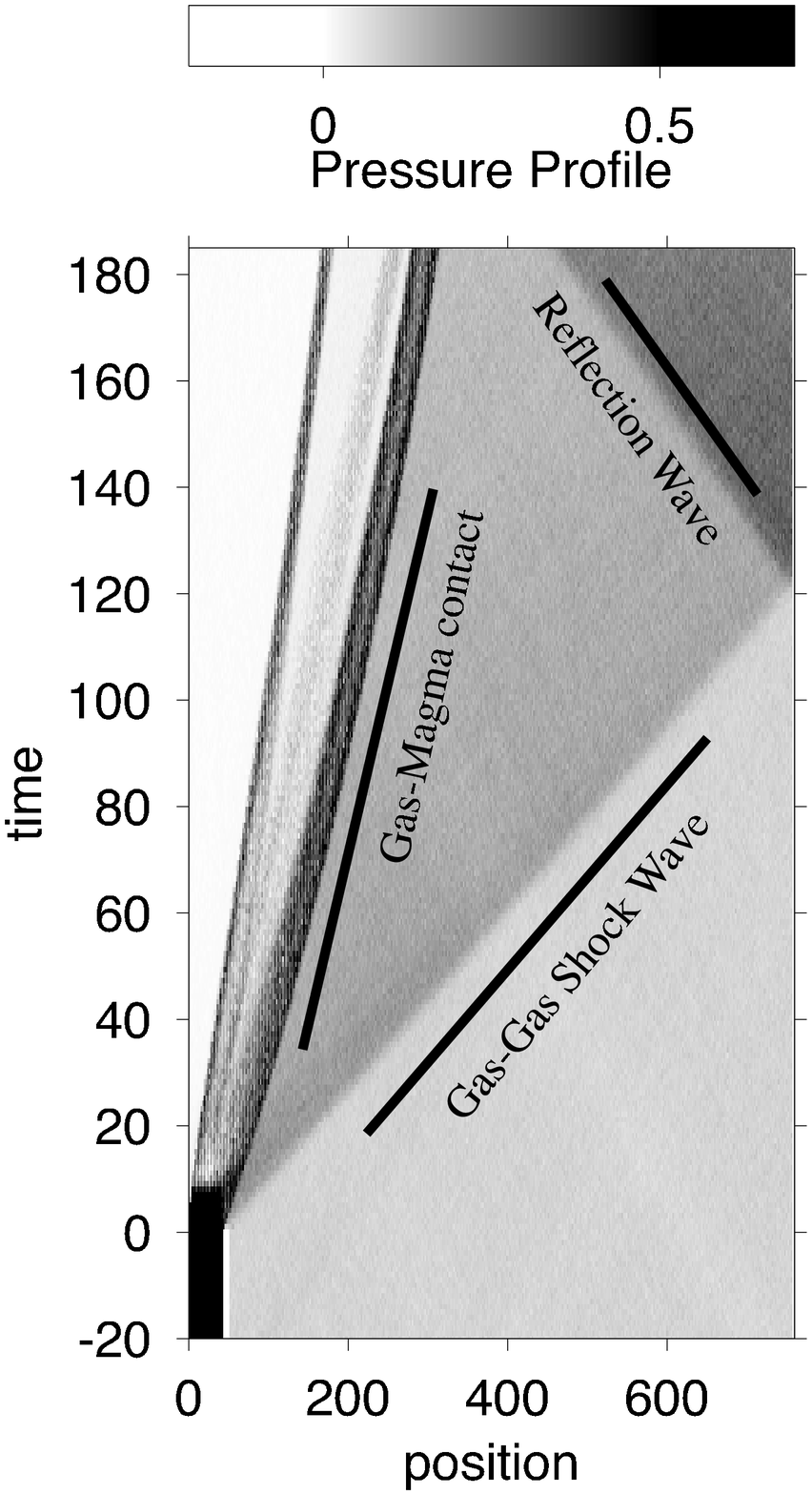} 
 \caption{Space-time profile of number density (left) and local
  pressure (right): Horizontal axis represents coordinate of explosion
  direction ($z$ axis) and vertical axis is time. At time $0$,
  a diaphragm is removed. Characteristic waves are guided by lines.
}
  \label{fig:spacetime}
\end{figure}
Here we present a typical result of simulation as space-time profile
of physical quantities.
In Figs.~\ref{fig:spacetime}, number density $n(z)$ and pressure  $p(z)$ are presented. 
In this
simulation, we take following parameters: System size is $L_x =L_y = 40, L_z
= 740. $ 
In an initial thermal equilibration stage, a diaphragm is located at
$z=40$, so the size of magma chamber is $40 \times
40 \times 40$ and one of the conduit is $40 \times
40 \times 700.$ 
Total number of particle is $176\,000$, which consists
of $57\,600$ magma particles and $118\,400$ gas particles. 
The chamber contains $57\,600$ magma
particles and $6\,400$ gas particles.
Other $112\,000$ gas particles are
in the conduit. 
Then initial number densities are $1$ for the chamber and $0.1$ for the conduit. 
Thermalization is done with the chamber temperature $2
$ and the conduit temperature $0.8$.

In Figs.~\ref{fig:spacetime}, a horizontal axis corresponds to $z$
direction and explosion goes to right. A vertical axis represents
time. At the time $0$, the diaphragm is removed. 
In the profile of number density, we recognize two characteristic
density waves. First one begins at $(z=40,t=0)$ and propagates to
$(750,120)$. This wave corresponds to a shock wave between hot
gas, which is heated by adiabatic compressing, 
and thermal equilibrium gas. Its velocity is larger than a sound
velocity of equilibrium conduit gas. This wave is reflected at
$(757,120)$, because an elastic wall exists at there.
Another wave propagates more
slowly than the shock wave from $(40,0)$ to $(300,185)$. 
Front position of this density wave corresponds to magma-gas contact
surface.

There are other small waves in this figure. A wave propagating 
from $(0,10)$ to $(170,185)$ is also reflecting wave caused by
the elastic wall located at $z=0$. A wave propagating to opposite
direction, which is from
$(40,0)$ to $(0,10)$,  is also observed in the figure. This
wave is an expansion wave of dense magma-gas mixture.

\begin{figure}[t]
  \centering
 \includegraphics[width=0.45 \textwidth]{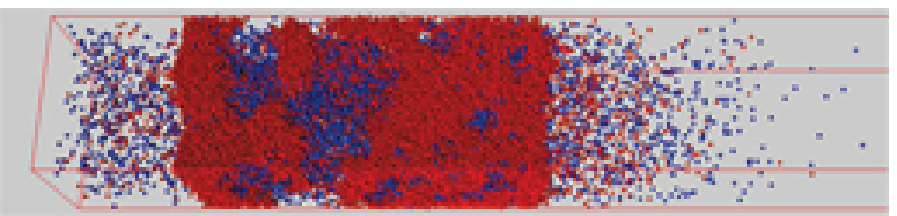}
  \centering
\includegraphics[width=0.45 \textwidth]{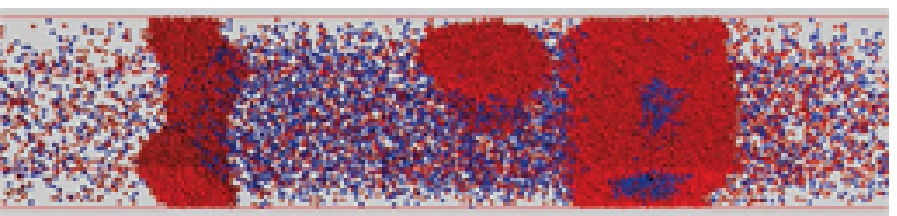}
  \caption{(Color online) Snapshots of simulation: (Up) Snapshot at $t=40$. (Down)
    Snapshot at $t=170$. Parameters are identical to ones
    of Fig.~\ref{fig:spacetime}. Eruption propagates to the right
    direction. Only particles originated from the chamber are
    plotted; A red ball represents a magma particle, and blue one is
    a gas particle. At the initial condition $t=0$, blue and red
    particles are uniformly mixed in the chamber.
}
  \label{fig:snapshot}
\end{figure}
Other significant features are observed in this space-time profile. 
After propagating magma-gas contact wave, some internal structures are
glowing. To investigate the internal structure in details, we show 
snapshot of simulation are shown in Figs.~\ref{fig:snapshot}. 
These
snapshot are taken from the simulation drawing
Fig.~\ref{fig:spacetime}, so simulation parameters are identical ones
of that simulation. We only draw magma particles and gas particles
which are in the magma chamber at the initial condition. Gas particles
coming from the conduit are omitted.  Explosion propagates to the
right direction in this figure, which is $z$ axis.

Before removing the diaphragm, magma and gas are uniformly mixed
in the magma chamber. But, in Figs.~\ref{fig:snapshot}, 
inhomogeneous mixing of 
those components is gradually growing
during the eruption. 
This reminds us of spinodal decomposition.
Size of exsolved gas bubble grows from 
Figs.~\ref{fig:snapshot}(a) to (b); In Fig.~\ref{fig:snapshot}(a),
bubble size are widely distributed but, in (b), one large  gas bubble 
and small bubbles in the thick magma exists. 
In large gas bubble, one magma droplet is observed.

In this way, magma-gas mixture become inhomogeneous mixture and
internal structure of bubbles are growing. Such behavior is consistent
with the scenario of volcanology. But in the present simulation,
transition to magma dispersion flow is not observed.  The reason may
be that smaller cross section of conduit and finiteness particles.

Next we compare the present simulation results with 
the continuum description given by Woods.\cite{Wo95}
In his model, magma-gas mixture is described by 
one-dimensional nonviscotic compressible one-component fluid.  
The dynamics are described by 
a continuity equation, an equation of motion, and 
the followings: 
\begin{align}
 \dfrac{1-n}{\rho_l} + \dfrac{nRT}{p_g} 
& = \dfrac{1}{\rho}, & 
p_g \left(\dfrac{\phi}{\rho} \right)^{\gamma_m} & = \mathrm{const}. 
\enspace, 
\end{align}
where $\rho, \rho_l, p_g,T,R,n,\phi$ and $\gamma_m$ denote 
mass density, 
mass density of magma component,
pressure of gas component,
temperature,
a gas constant,
a mass fraction of magma and gas components,
a volume fraction of magma and gas components,
and ratio of specific heats,
respectively.  
In these quantities, $\rho, p_g, T$  and $\phi^{-1} \equiv 1+\frac{1-n}{n} \frac{p_g}{\rho_l R T} $ are variables.
Other $\rho_l, R, n, $ and $\gamma_m$ are  fixed to some constant values. 
The first equation is an equation of states, and the second one 
expresses an isentropic condition derived from the first law of thermodynamics.
As we know the present equation of states is almost identical to one of ideal gas. 

These equations are essentially same as ones of compressible ideal gas fluid.
To study the equations is just an textbook example.\cite{La49,La59}
We get a standard rewrite as 
\begin{equation}
\left\{
\dfrac{\partial\,\,}{\partial t} 
+ \left( w \pm a(\rho) \right) \dfrac{\partial \,\,}{\partial z} 
\right\} 
\left( 
w \pm \int^\rho \dfrac{a(\rho')}{\rho'}d\rho'
\right) = 0 \enspace ,
\label{eq:standard}
\end{equation}
where $w$ and $a(\rho)$ are a velocity field and a sound velocity, respectively.
The sound velocity of magma-gas mixture is a
function of $\rho$, and it is expressed as 
$a^2(\rho) = a^2_0 (\rho/\rho_0)^{\gamma_m-1}( \phi_0/\phi)^{\gamma_m+1}$
($a_0,\rho_0,\phi_0$ are sound velocity, density, and volume fraction
at some reference state.)
This equation gives characteristic curves and conserved quantities
on them. Then we can solve the equation in characteristic regions. 
For obtaining global shock tube solution, we have to
glue the solution with appropriate boundary conditions.

\begin{figure}[htbp]
\centering
\includegraphics[width=0.4\textwidth]{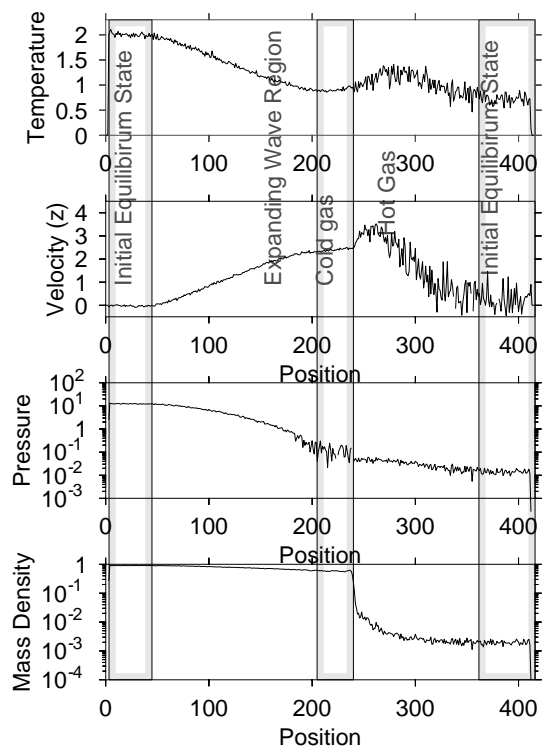}
\caption{Spatial profiles of temperature, velocity ($z$), pressure,
  and mass density at $t=15$:
  System size is taken to be $L_x = 32, L_y = 32, L_z = 408$ and
  size of magma chamber is $ 32 \times 32\times 200$.
 Initial mass density and temperature are taken to be $1$ and
  $2$, respectively.
  We can recognize characteristic
  regions. From right, ``initial equilibrium state'', ``hot gas
  region'', ``cold gas region'', ``expanding wave  region'', and
  ``initial equilibrium state'' again are observed. These regions are
  indicated by gray rectangular. 
} 
\label{fig:prof15}
\end{figure}
In Fig.~\ref{fig:prof15}, 
temperature $T(z)$, barycentric velocity $\mathbf{v}(z)_z$,
pressure $p(z)$, mass density $\rho(z)$ of the present simulation are shown. 
Simulation parameters are taken to be as follows: System size is $L_x
= L_y = 32, L_z = 408$ and size of magma chamber is $32 \times 32
\times 200$. Initial number density of magma chamber is taken to be
$1$ and conduit density is $0.02$, thus the number of particles in the
chamber is $204\,800$, which contains $10$\% gas particles. The
number of gas particles in the conduit is $4\,096$.  In this simulation, we
imposed an artificial boundary condition at the top of conduit; For
decreasing reflection effects from the top elastic wall, we attach a particle
sink at the top, in which particles with the energy larger than some
threshold value are removed from the system.

We can observe characteristic regions in Fig.~\ref{fig:prof15}. 
Let us compare these results with continuum descriptions. 
The solution of Eq.~(\ref{eq:standard}) 
teaches us that there are three regions in shock tube
analysis, that is, a hot gas region, a cold gas region, and an expanding wave region.
Corresponding regions of simulation are indicated in the figure;
In the ``hot gas'' region, gases are heating up by the shock wave. 
In contrast, in the ``cold gas'' region, 
gases are cooling by an adiabatic expansion. 
Another region is  an ``expanding wave'' region in which the expanding wave exists 
and physical quantities are smoothly changed. 
Physical properties of such regions obtained by the simulation 
are almost equivalent to ones of shock tube analysis. 
But there is a little mismatch with the solution; 
Analysis of compressible 
fluid gives constant profiles of physical quantities in both hot and
cold gas regions.  But, in this simulation, some structures are
observed in each regions.  For example, in a velocity profile of the
hot gas region, velocity near cold gas is rather
faster than other areas. This high velocity area is caused by
pushing effects of magma-gas contact surface, which is corresponding
to the front of cold gas contact. These high velocity particles are
not thermalized yet; In the molecular dynamics simulation, microscopic
relaxation is apparently observed.



To summarize, we have constructed a microscopic model of Vulcanian
eruption by a two-components Lennard-Jones particle system. 
We observed that the particle dynamics is efficient in this kind of dynamics. 
Using the present model we can reproduce characteristic features of 
explosive eruption such as a shock wave, a expansion wave. 
At the early stage of the eruption, we also compare the simulation
result with the analytic model given by Woods. Qualitative behavior is 
almost consistent with the analytic result, 
even though the flow is treated as nonviscotic one in the analytic model. 
In addition, we have also observed that the internal structure is
growing during the eruption.
Internal bubble structure cannot be captured by the Woods model. 
This behavior is also consistent with a eruption picture of volcanology study. 
Thus we conclude that the present model is a candidate of 
``an Ising model of Vulcanian eruption''.

To establish the present model, a quantitative study is inevitable. 
For this purpose, we have to enlarge the size of system; 
A transition from bubbly magma flow to magma dispersion flow will be reproduced 
and studied by simulation of the system with ten times larger to all directions. And the more
details of volcanic eruption not only Vulcanian but also Strombolian, and Plinian will be elucidated. 
Present typical computational time is approximately 80 hours for $208\,896$ particles 
with single AMD opteron 248 (2.2GHz). Hence much larger simulation is feasible  with large super computers. 


\section*{Acknowledgments}

The authors thank T. Koyaguchi for valuable discussion and comments.
This work is partially supported by the Ministry of Education, Science, Sports and Culture,
Grant-in-Aid for Scientific Research Priority Areas, No.14080204, 2005.
The part of computation in this work has been done using the facilities of the Supercomputer Center, Institute for Solid State Physics, University of Tokyo and the Earth Simulator Center, Japan Agency for Marine-Earth Science and Technology.

\end{document}